\documentclass[runningheads]{llncs}

\usepackage{graphicx}
\usepackage{color}
\usepackage{microtype}
\usepackage{amsmath}
\usepackage{graphicx}
\usepackage{xcolor}
\usepackage{hyperref}
\usepackage{algorithm}
\usepackage{listings}
\usepackage{tabularx}
\usepackage{soul}
\usepackage{cleveref}
\usepackage{url}
\usepackage{comment}
\usepackage{fancyvrb}
\usepackage{svg}
\usepackage{verbatimbox}
\usepackage[T1]{fontenc}
\usepackage{subcaption}
\usepackage{tcolorbox}
\usepackage{lmodern} % for bold teletype font
\usepackage[frozencache,cachedir=.]{minted}
\begin{document}
%
%\title{Contribution Title\thanks{Supported by organization x.}}
\title{ESBMC v7.3: Model Checking C++ Programs using Clang AST}
%
%\titlerunning{Abbreviated paper title}
% If the paper title is too long for the running head, you can set
% an abbreviated paper title here
%
\author{
  Kunjian Song\inst{1}
\and
  Mikhail R. Gadelha\inst{2}
\and
  Franz Brauße\inst{1}
\and
  Rafael S. Menezes\inst{1}
\and
  Lucas C. Cordeiro\inst{1}
}
\authorrunning{K. Song et al.}
% First names are abbreviated in the running head.
% If there are more than two authors, 'et al.' is used.
%
\institute{
  The University of Manchester, UK \\
  \email{\{kunjian.song, rafael.menezes\}@postgrad.manchester.ac.uk,\\
  \{franz.brausse, lucas.cordeiro\}@manchester.ac.uk} \\
\and
  Igalia, A Coruña, Spain \\
  \email{mikhail@igalia.com} \\
}
% reset footnote number to zero for the actual paper content
\let\oldmaketitle\maketitle
\renewcommand{\maketitle}{\oldmaketitle\setcounter{footnote}{0}}
\maketitle              % typeset the header of the contribution

%
%The abstract should briefly summarize the contents of the paper in
%15--250 words.
\begin{abstract}
This paper introduces ESBMC v7.3, the latest Efficient SMT-Based Context-Bounded Model Checker version, which now incorporates a new clang-based C++ front-end. While the previous CPROVER-based front-end served well for handling C++03 programs, it encountered challenges keeping up with the evolving C++ language. As new language and library features were added in each C++ version, the limitations of the old front-end became apparent, leading to difficult-to-maintain code. Consequently, modern C++ programs were challenging to verify. To overcome this obstacle, we redeveloped the front-end, opting for a more robust approach using clang. The new front-end efficiently traverses the Abstract Syntax Tree (AST) in-memory using clang APIs and transforms each AST node into ESBMC's Intermediate Representation. Through extensive experimentation, our results demonstrate that ESBMC v7.3 with the new front-end significantly reduces parse and conversion errors, enabling successful verification of a wide range of C++ programs, thereby outperforming previous ESBMC versions.

\end{abstract}
%
%
%
%====================================================================
\section{Introduction}
\label{intro}
%====================================================================

C++ is one of the most popular programming languages used to build high-performance and real-time systems, such as operating systems, banking systems, communication systems, and embedded systems~\cite{deitel2014c++,CordeiroFB20}. However, memory safety issues remain a major source of security vulnerabilities in C++ programs~\cite{miller2019trends}. Fan et al.~\cite{fan2020ac} created a dataset of C/C++ vulnerabilities by mining the Common Vulnerabilities and Exposures (CVE) database~\cite{cvemitreurl} and the associated open-source projects on GitHub, then curated the issues based on Common Weakness Enumeration (CWE)~\cite{cwemitreurl}. According to their findings, two out of the top three vulnerabilities are caused by memory safety issues: Improper Restriction of Operations within the Bounds of a Memory Buffer (CWE-119) and Out-of-bounds Read (CWE-125)~\cite{fan2020ac}. 

The limitation of software testing resides in the user inputs~\cite{quadri2010software}. Only a limited number of execution paths may be tested since test cases involve human inputs in the form of concrete values~\cite{ammann2016introduction}. Unlike testing, formal verification techniques can be used more systematically to reason about a program, although they suffer from the state-space explosion problem~\cite{MonteiroGCF17}. There is an increasing adoption of formal verification techniques for C programs in the industry, e.g., Amazon has been using model-checking techniques to prove the correctness of their C-based systems in Amazon Web Services (AWS); this has positively impacted their code quality, as evidenced by the increased rate of bugs found and fixed~\cite{chong2020code}. 

Formal verification of C++ programs is more challenging than C programs due to the sophisticated features, such as the STL (Standard Template Libraries) containers, templates, exception handling, and object-oriented programming (OOP) paradigm~\cite{deitel2014c++}. The existing state-of-the-art verification tools for C++ programs only have limited feature support~\cite{monteiro2022model}. For ESBMC, Ramalho et al.~\cite{ramalho2013smt} and Monteiro et al.~\cite{monteiro2022model} initiated the support for C++ program verification. Since then, ESBMC has undergone heavy development. 

This research presents a significant improvement to ESBMC's C++ verification capabilities by introducing a new clang-based frontend. Particularly, the original contributions of this work are as follows:
\begin{itemize}
\item \textbf{Complete Redesign}: ESBMC's C++ frontend has undergone a complete overhaul and now relies on clang~\cite{clangllvm}. By leveraging Clang's parsing and semantic analysis capabilities~\cite{lopes2014getting,pandey2015llvm}, we check the input program's Abstract Syntax Tree (AST) using a production-quality compiler. This eliminates the need for static analysis logic and ensures enhanced accuracy and efficiency.
\item \textbf{Object Models Details}: We provide comprehensive insights into the object models used to achieve seamless conversion of C++ polymorphism code to ESBMC's Intermediate Representation (IR). This improvement allows ESBMC to handle C++ growth and its variants like CUDA~\cite{PereiraASMMFC17}.
\item \textbf{Simplified Type Checking for Templates:} The new clang-based frontend greatly simplifies type checking for templates, streamlining ESBMC's ability to adapt to C++ advancements. Furthermore, this enhancement facilitates the incorporation of C++ variants like CUDA.
\end{itemize}

By introducing these advancements, our work significantly enhances ESBMC's C++ verification capabilities, paving the way for more robust and efficient verification of C++ programs and their variants.

%====================================================================
\section{Background}
\label{background}
%====================================================================

ESBMC's verification for C++03 programs reaches its maturity in version v2.1, presented by Monteiro et al.~\cite{monteiro2022model}. ESBMC v2.1 provides a first-order logic-based framework that formalizes a wide range of C++ core languages, verifying the input C++ programs by encoding them into SMT formulas. Since C++ Standard Template Libraries (STL) contain optimized assembly code not verifiable using ESBMC, ESBMC v2.1 tackled this problem using a collection of C++ operational models (OM) to replace the STL included in the input program. The OMs are abstract representations mimicking the structure of the STL, adding pre- and post-conditions to all STL APIs~\cite{dos2015simple}. Combining these approaches, ESBMC v2.1 outperformed other state-of-the-art tools evaluated over a large set of benchmarks, comprising $1513$ test cases~\cite{monteiro2022model}. Nonetheless, ESBMC v2.1 employs a Flex and Bison-based frontend from CBMC~\cite{clarke2004tool}, which leads to hard-to-main code and can hardly evolve to support modern C++11 features. 

%---------------------------------------------------------------
\subsubsection{Limitations of the old C++ frontend}
\label{limitations-of-old-frontend}
%---------------------------------------------------------------

The version of ESBMC in Monteiro et al.~\cite{monteiro2022model} uses an outdated CPROVER-based frontend~\cite{clarke2004tool} with the following limitations.

\begin{enumerate}
  \item For the type-checking phase, ESBMC could not provide meaningful warnings or error messages.
  \item It is inefficient at generating a body for default implicit non-trivial methods in a class, such as C++ copy constructors or copy assignment operators 
  \item The parser of the old front-end needs to be manually updated to cover the essential C++ semantic rules~\cite{esbmccpptypecheck}, which leads to hard-to-maintain code to keep up with the C++ evolution. 
  \item The old front-end contains excessive data structures and procedures auxiliary to scope resolution and function type checking.
  \item The type checker~\cite{esbmccpptypecheck} of the old frontend only works with a CPROVER-based parse tree and supports up to C++03 standard~\cite{cpp03std}. We find adapting it to the new C++ language and library features difficult. 
  \item The old front-end uses a speculative approach to guess the arguments for a template specialization and a map to associate the template parameters to their instantiated values, which leads to hard-to-maintain and hard-to-debug code in the case of recursive templates. Additionally, owing to its limited static analysis, the old front-end could not provide any early warning when there is a circular dependency on the templates.
\end{enumerate}

These limitations combine to a point where the old front-end is too laborious to maintain and extend for formal verification of modern C++ programs. We propose the clang-based approach to convert an input C++ program to ESBMC's IR to overcome these limitations. 

%====================================================================
\section{Model Checking C++ Programs using Clang AST}
\label{clang-based approach}
%====================================================================

\begin{figure*}[h]
  \centering
  \includegraphics[width=1\textwidth]{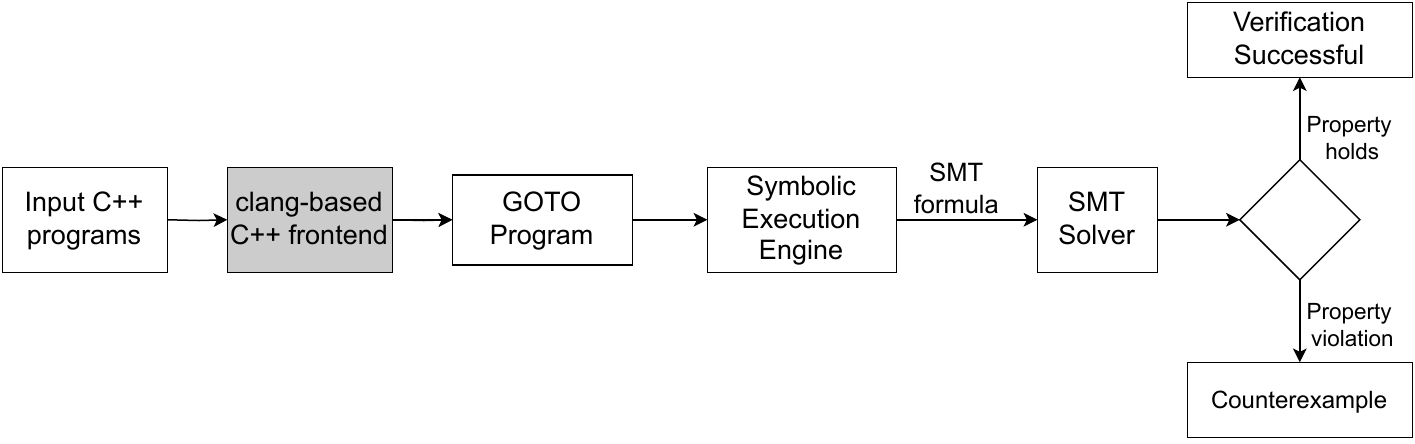}
  \vspace{-1.2em}
  \caption{ESBMC architecture for C++ verification. The grey block represents the new clang-based C++ front-end integrated into ESBMC v7.3.}
  \label{fig:esbmcpp-architecture}
\end{figure*}

Figure~\ref{fig:esbmcpp-architecture} illustrates ESBMC's verification pipeline for C++ programs. The new clang-cpp frontend type-checks and converts the input C++ program (along with the corresponding OMs) into the GOTO program representation~\cite{cordeiro2011smt,CordeiroF11}. Then the GOTO program will be symbolically executed to generate the SSA form of the program, thus generating a set of logical formulas consisting of the constraints and properties. An SMT solver is used to check the satisfiability of the formulas, giving a verdict \textit{VERFICATION SUCCESSFUL} if no property violation is found up the bound $k$ or a counterexample in case of property violation. 

%-------------------------------------------------
\subsection{Polymorphism}
\label{polymorphism}
%-------------------------------------------------

The traditional approach for achieving polymorphism makes use of virtual function tables (also known as \textit{vtables}) and virtual pointers (known as \textit{vptrs}). While the clang AST, to the best of our knowledge, does not include information about virtual tables or virtual pointers of a class, it nonetheless provides users with enough information to enable them to create their \textit{vtables} and \textit{vptrs}. In the new clang-based C++ frontend, we reimplemented the \textit{vtable} and \textit{vptr} construction mechanism following a similar approach from ESBMC v2.1, but with significant simplifications based on the information provided in the clang AST. Figure~\ref{fig:cpp_poly} illustrates an example of C++ polymorphism.

\begin{figure}[ht]
  \centering
    \begin{subfigure}{.5\textwidth}
    \begin{minted}[xleftmargin=20pt,linenos]{cpp}
class Bird {
  public:
  virtual int doit(void) { return 21; }
};

class Penguin: public Bird {
  public:
  int doit(void) override { return 42; }
};
int main(){
  Bird *p = new Penguin();
  assert(p->doit() == 42);
  delete p;
  return 0;
}
    \end{minted}
    \end{subfigure}
\caption{Example of C++ classes with virtual functions.}
\label{fig:cpp_poly}
\end{figure}

\begin{figure}[!ht]
  \centering
  \includegraphics[width=1\textwidth]{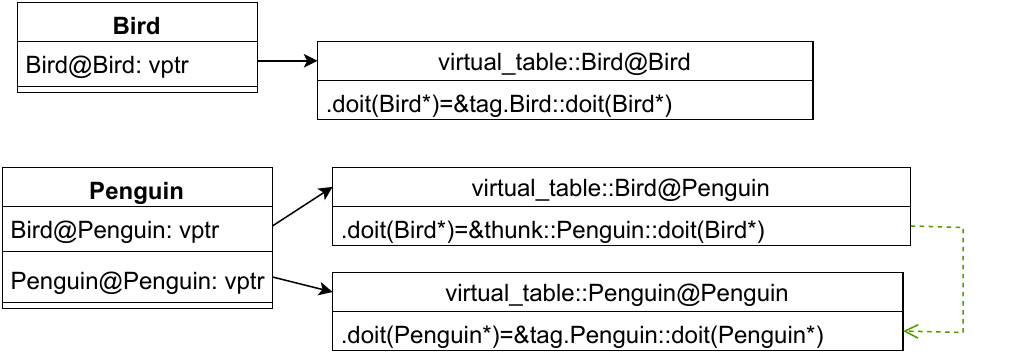}
  \vspace{-1.2em}
  \caption{Object models for Bird and Penguin classes}
  \label{fig:cpp_poly_objmodel}
\end{figure}

Figure~\ref{fig:cpp_poly_objmodel} illustrates the object models for the Bird and Penguin classes. The new front-end adds one or more \textit{vptrs} to each class. The \textit{vptrs} will be initialized in the class constructors, which set each \textit{vptr} pointing to the desired \textit{vtable}. The child class contains an additional pointer pointing to a \textit{vtable} with a thunk to the overriding function. The thunk redirects the call to the corresponding overriding function. In the case of multiple inheritances, the child class would have multiple \textit{vtprs} ``inherited'' from multiple base classes. The new front-end can also manage a virtual inheritance, such as the diamond problem, which avoids duplicating \textit{vptrs}, referring to the same virtual table in an inheritance hierarchy. Line 2-4 in Figure~\ref{fig:cpp_poly_goto_sub1} illustrates the dynamic dispatch is achieved using the \textit{vptr} calling the thunk, which in turn calls the desired overriding function in Figure~\ref{fig:cpp_poly_goto_sub2} Line 9-11. Note that the \textit{override} specifier is a C++11 extension that the old front-end could not support. 

\begin{figure*}[!h]
  \centering
    \begin{subfigure}[t]{.4\textwidth}
    \begin{lstlisting}[numbers=left,label={lst:unions},mathescape,language=c,tabsize=2
     ,morekeywords={signed int, assert, ->}]
int return_value;
return_value =
*p->Bird@Penguin
    ->doit(p)
assert(return_value == 42)
    \end{lstlisting}
    \caption{GOTO program of the dynamic dispatch in Line 12 of Figure~\ref{fig:cpp_poly}.}
    \label{fig:cpp_poly_goto_sub1}
    \end{subfigure}
    \hspace{1.cm}
    \begin{subfigure}[t]{.4\textwidth}
    \begin{lstlisting}[numbers=left,label={lst:unions2},mathescape,language=c
     ,tabsize=2,morekeywords={signed int, assert, ->}]
thunk::Penguin::doit(Bird*):
  int return_value;
  return_value =
    Penguin::doit(
      (Penguin*)this)
  RETURN: return_value
  END_FUNCTION

Penguin::doit(Penguin*):
  RETURN: 42
  END_FUNCTION
    \end{lstlisting}
    \caption{thunk redirecting the call to the overriding function.}
    \label{fig:cpp_poly_goto_sub2}
    \end{subfigure}
\caption{GOTO conversions of the overriding methods and dynamic dispatch.}
\label{fig:cpp_poly_goto}
\end{figure*}

%------------------------------------------
\subsection{Template}
\label{template} 
%------------------------------------------

Template is a key feature in C++, allowing type to be passed as a parameter. The template allows STL containers and generic algorithms to work with different C++ data types~\cite{prata2012c++,stroustrup2013c++}. The old front-end in ESBMC v2.1 implements the template specialization based on Siek et al.~\cite{siek2006semantic,monteiro2022model}. However, it produces a ``CONVERSION ERROR'' for the test case illustrated in Figure~\ref{fig:template_example_code}. This benchmark is based on the \textit{Friend18} example from the GCC test suite~\cite{gcctestsuite}, which was added for Bug $10158$ on GCC Bugzilla~\cite{gcc10158}. ESBMC v7.3 successfully verified this benchmark and found the assertion's property violation in Figure~\ref{fig:template_example_code}. The verification result is illustrated in Figure~\ref{fig:tempalte_verfication_verdict}. The example in Figure~\ref{fig:template_example_code} contains a C++20 extension. The \textit{foo} function is defined in \textit{struct X}, but gets called using an unqualified name with explicit template arguments in \textit{main}. ESBMC v2.1 failed to verify it due to the ``CONVERSION ERROR symbol ```foo' not found''. We also tried this example with CBMC 5.88.1~\cite{cbmc5881rel}, which aborted during type-checking, and cppcheck v2.11.1~\cite{cppcheck2111rel}, which did not give any verification verdict. 

\begin{figure*}[!h]
  \centering
    \begin{subfigure}[t]{.4\textwidth}
    \begin{minted}[xleftmargin=20pt,linenos]{cpp}
#include <cassert>
template <int N> struct X
{
  template <int M>
  friend int foo(X const &)
  {
    return N * 10000 + M;
  }
};
X<1234> bring;

int main() {
  assert(
   foo<5678> (bring)
    !=12345678);
}
    \end{minted}
    \caption{Example of C++ class template}
    \label{fig:template_example_code}
    \end{subfigure}
    \hspace{1.cm}
    \begin{subfigure}[t]{.4\textwidth}
    \begin{lstlisting}[numbers=left,label={lst:unions2},mathescape,language=c
     ,tabsize=2,morekeywords={signed int, assert, ->}]
Violated property:
  file tmp2.cpp 
   line 13 column 3 
    function main
  assertion 
   foo<5678>(bring)!=12345678
  return_value!=12345678

VERIFICATION FAILED
    \end{lstlisting}
    \caption{Verdict for the template example}
    \label{fig:tempalte_verfication_verdict}
    \end{subfigure}
\caption{ESBMC verified the \textit{Friend18} example from the GCC test suite.~\cite{gcctestsuite}}
\label{fig:cpp_template_example}
\end{figure*}

%====================================================================
\section{Experimental Evaluation}
\label{experimental-evaluation}
%====================================================================

We used some benchmarks from Monteiro et al.~\cite{monteiro2022model} to evaluate ESBMC v7.3. These benchmarks were used to assess ESBMC v2.1. Still, we excluded the other two verifiers (LLBMC~\cite{llbmcmodelchker} and DIVINE~\cite{divinemirrorgithub}) for the following reasons: (1) they have been already evaluated by Monteiro et al.~\cite{monteiro2022model} and ESBMC v2.1 was found outperforming them; (2) to our best knowledge and effort, we were unable to get a working version of them. The last version of LLBMC was released in $2013$, and its download link is currently broken. The last commit to DIVINE's mirror repository on GitHub dated back to Mar $2021$, and DIVINE failed to build.  

We did not evaluate the test cases (TCs) that depend on the operational models (OMs) in each benchmark. We only ran the TCs for core C++ language features because the OMs for the new clang-based C++ front-end are still under development, e.g., exception handling support. Otherwise, running test cases for sure to fail would be pointless due to a feature still being developed. Hence each benchmark is a subset of the original benchmark, which only comprises TCs for verifying core C++ language features. There are $399$ benchmarks in total over $6$ sub-benchmarks. The \textit{cpp-sub} contains example programs from the book \textit{C++ How to Program}~\cite{deitel2007how}. The inheritance and polymorphism sub-benchmarks are extracted from~\cite{monteiro2022model}. There are three sub-benchmarks for template specialization - \textit{cbmc-sub} comes from the CBMC regressions~\cite{cbmctestsuite}; \textit{gcc-template-tests-sub} were extracted from the GCC template test suite~\cite{gcctestsuite}; \textit{template-sub} is also from benchmarks used in~\cite{monteiro2022model}. \textit{cpp-sub} contains programs with mixed use of various C++ language features combined with inheritance, polymorphism, and templates.

%-----------------------------------------------
\subsection{Objectives and Setup}
\label{objectives} 
%-----------------------------------------------

Our evaluation framework is based on Python's \textit{unittest}~\cite{pyunittest}. For each TC in the test suite, we check whether the verification verdict reported by each tool matches the expected outcome. TC passes when the tool reports a verdict of ``VERIFICATION SUCCESSFUL'' on a program without any violation of properties or reports ``VERIFICATION FAILED'' on an unsafe program that violates a property. Such properties include arithmetic overflows, array out-of-bounds, memory issues, or assertion failures. Our evaluation aims to answer the following experimental questions: 
\begin{enumerate}
\item[\textbf{EQ1}]: (\textbf{soundness}) Can ESBMC give more correct verification results and a higher pass rate than its previous versions?
\item[\textbf{EQ2}]: (\textbf{performance}) How long does ESBMC v7.3 take to verify C++ programs? 
\item[\textbf{EQ3}]: (\textbf{completeness}) Does the tool complete the future work specified by Monteiro et al.~\cite{monteiro2022model}?
\end{enumerate}

The experiment was set up in Ubuntu 20.04 with $32$GB RAM on an $8$-core Intel CPU. The dataset, scripts, and logs are publicly available in Zenodo~\cite{zenodoarchive}. The accumulative verification time represents the CPU time elapsed for each tool finishing all sub-benchmarks. 

%-------------------------------
\subsection{Results}
\label{results} 
%-------------------------------

Table~\ref{table:Evaluation-results} shows our experimental results. With a higher pass rate than ESBMC v2.1 over $5$ out of $6$ sub-benchmarks, ESBMC v7.3 successfully verified all benchmarks and passed all test cases, confirming \textbf{EQ1}. As for ESBMC v2.1, the failed TCs in \textit{cpp-sub} are due to parsing or conversion errors, meaning the previous tool version is unable to properly type-check the input programs, probably due to the weak parser, as described in Section~\ref{background}. The failed TCs in \textit{inheritance and polymorphism-sub} contain a common feature of dynamically casting a pointer of a child class with a base class containing virtual methods. ESBMC v2.1 could not handle this type of casting, giving conversion errors. 

ESBMC v2.1 has limited support for C++ templates, matching our expectations as reported by Monteiro et al.~\cite{monteiro2022model}. The failed test cases in \textit{cbmc-template-sub} are the results of ESBMC v2.1 not able to handle the default template type parameter or explicit template specialization combined with C++ \textit{typedef} specifier. The low pass rate of ESBMC v2.1 on \textit{gcc-template-tests-sub} indicates that the old version cannot verify test cases used by an industrial compiler. \textbf{EQ3} is affirmed through the experiment, as none of these problems persist in ESBMC v7.3. Since one of the test cases in \textit{cpp-sub} timed out against ESBMC v2.1 after $900$ seconds, the actual verification time has been rectified to $149$s; otherwise, the cumulative verification time would be $1049$s. As for the performance \textbf{EQ2}, ESBMC v7.3 could verify all sub-benchmarks in $128$s, faster than its previous version, which affirms \textbf{EQ2}. 
\begin{table}[h]
\caption{Experimental results showing the pass rate for each sub-benchmark and accumulative verification time.}
\begin{center} 
\begin{tabular}{|c|c|c|c|c|c|c|c|c|c|c|c|}
\hline % \cline{i-j}, line from column i to column j
\textbf{Sub-Benchmarks} &\textbf{ESBMC-v2.1 pass rate} &\textbf{ESBMC-v7.3 pass rate} \\ \hline
    cpp-sub                 &{$91\%$}       &{$100\%$}     \\\hline
    inheritance-sub         &{$79\%$}       &{$100\%$}     \\\hline
    polymorphism-sub        &{$87\%$}       &{$100\%$}     \\\hline
    cbmc-template-sub       &{$92\%$}       &{$100\%$}     \\\hline
    gcc-template-tests-sub  &{$39\%$}       &{$100\%$}     \\\hline
    template-sub            &{$100\%$}      &{$100\%$}     \\\hline
    
\textbf{Total verification Time}     &{$149.94s$} &{$128.796s$}    \\\hline

\end{tabular}
\label{table:Evaluation-results}
\end{center}
\end{table}

% answers to EQ1, 2 and 3:
Overall, we have enhanced the template support in ESBMC v7.3, which completed the future work by Monteiro et al.~\cite{monteiro2022model}. In comparison to its previous version, ESBMC v7.3 can provide more accurate results faster. 

%-------------------------------------------
\subsection{Threats to Validity}
\label{threats-to-validity}
%-------------------------------------------

While developing the new C++ frontend, we found that the clang AST does not fully describe the correct order of constructors or destructors to be called in the most derived class in a complex hierarchical inheritance graph, e.g., crossed diamond hierarchy. We documented it under an umbrella issue, which is currently in our backlog~\cite{esbmccppsupport} on ESBMC GitHub repository~\cite{esbmcissue940}. We might need to use an additional data structure to keep track of the most derived class and implement an algorithm to recursively describe the correct order of base initialization or destruction in the class inheritance graph, which remains an open challenge.

%====================================================================
\section{Conclusion and Future Work}
\label{conclusion-and-future-work}
%====================================================================

We present a new clang-based front-end that converts in-memory clang AST to ESBMC's IR. In our evaluation of ESBMC v7.3, we compared it to ESBMC v2.1, specifically focusing on a subset of benchmarks to cover core C++ language features. The results demonstrate significant progress with ESBMC v7.3, as it successfully parses real-world C++ programs, including those from the GCC test suite. Notably, it significantly reduces the number of conversion and parse errors compared to the previous version, showcasing improved performance over the sub-benchmarks for core language features. 

While ESBMC effectively mimics the semantics of APIs of the STL libraries using the OMs from ESBMC v2.1, we recognize the need for continuous improvement. As we endeavor to verify modern C++ programs, these OMs require regular review and updates to align with the C++ standard used in the input program. Accurate OMs are essential, as any approximation may lead to incorrect encoding and invalidate the verification results. To further enhance our front-end coverage and reduce the number of OMs we maintain, our future work will focus on handling more C++ libraries. 

Additionally, we aim to integrate various checkers, such as cppcheck~\cite{cppcheck2111rel}, into our testing framework to facilitate future evaluations. 
Our previous success verifying a commercial C++ telecommunication application using ESBMC v2.1 has inspired further goals~\cite{SousaCF15,monteiro2022model}. With ESBMC v7.3 and beyond, we plan to verify the C++ interpreter in OpenJDK as part of the Soteria project~\cite{soteriaorg} and contribute to benchmarks for the International Competition on Software Verification (SV-COMP)~\cite{Beyer23}. 

%These endeavors contribute to the continuous growth and enhancement of ESBMC, making it a powerful tool for verifying complex C++ programs and advancing the state-of-the-art in formal verification.

%
% ---- Bibliography ----
%
% BibTeX users should specify bibliography style 'splncs04'.
% References will then be sorted and formatted in the correct style.
%
% \bibliographystyle{splncs04}
% \bibliography{mybibliography}
%
\bibliographystyle{splncs04}
\bibliography{refs}

\appendix
%====================================================================
\section{Memory Consumption}
\label{memory-consumption}
%====================================================================
%\textcolor{red}{TODO: Add table for memory consumption}

In addition to the pass rate and verification time in Table~\ref{table:Evaluation-results}, we also assessed each tool's memory usage. Table~\ref{table:memory-usage} shows the cumulative maximum RSS (Resident Set Size) for each benchmark using each tool under evaluation. Our metrics collection approach is based on Python's $resource$ module~\footnote{\url{https://docs.python.org/3/library/resource.html}}, $subprocess$ module~\footnote{\url{https://docs.python.org/3/library/subprocess.html}} and \textit{unit test framework}~\cite{pyunittest}. 

%---------------------------------------------------------------
\subsubsection{Memory metrics collection approach}
\label{memory-metrics-collection-approach}
%---------------------------------------------------------------
The unit test framework encapsulates a benchmark in a test suite that launches a sub-process for each test case and waits for it to finish. We obtain the maximum RSS for each test case sub-process that has been terminated. In each row of Table~\ref{table:memory-usage}, the cumulative maximum RSS for a benchmark is calculated by summing the maximum RSS for each sub-process. The final row, ~\textit{Total memory}, totals the amount of memory used by each tool to perform each benchmark. 

\begin{table}[h]
\caption{Experimental results showing the cumulative maximum RSS (Resident Set Size) for each sub-benchmarks.}
\begin{center} 
\begin{tabular}{|c|c|c|c|c|c|c|c|c|c|c|c|}
\hline % \cline{i-j}, line from column i to column j
\textbf{Sub-Benchmarks} &\textbf{ESBMC-v2.1} &\textbf{ESBMC-v7.3} \\ \hline
    cpp-sub                 &{$31477$ MB}    &{$19385$ MB}  \\\hline
    inheritance-sub         &{$231$ MB}      &{$845$ MB}    \\\hline
    polymorphism-sub        &{$722$ MB}      &{$2373$ MB}   \\\hline
    cbmc-template-sub       &{$650$ MB}       &{$2295$ MB}   \\\hline
    gcc-template-tests-sub  &{$395$ MB}       &{$1387$ MB}   \\\hline
    template-sub            &{$207$ MB}       &{$727$ MB}    \\\hline
    
\textbf{Total memory} &{$33682$ MB} &{$27012$ MB}    \\\hline
\end{tabular}
\label{table:memory-usage}
\end{center}
\end{table}

Compared to ESBMC v2.1, ESBMC v7.3 can verify more test cases and uses less memory. The lower memory usage of v2.1 than v7.3 is due to lower pass rates for the benchmarks, mainly because of v2.1's inadequacy to handle C++ templates. Many TCs failed due to CONVERSION ERROR in ESBMC's front-end and never even reached the solver in ESBMC's backend. As a result, no verification effort was made for those TCs and hence less memory was used. 

%====================================================================
\section{Performance Using Different SMT Solvers}
\label{performance-smt-solvers}
%====================================================================
ESBMC supports multiple SMT solvers in the back-end, such as Z3~\cite{moura2008z3}, Bitwuzla~\cite{niemetz2020bitwuzla}, Boolector~\cite{brummayer2009boolector}, MathSAT~\cite{bruttomesso2008mathsat}, CVC4~\cite{barrett2011cvc4}, and Yices~\cite{dutertre2014yices}. We also evaluated ESBMC v7.3 with various solvers over the same set of benchmarks. Table~\ref{table:results-solvers-time} shows the pass rates and total verification time for ESBMC v7.3 using different solvers, and Table~\ref{table:results-solvers-memory} shows the memory consumption for the same experimental set-up using the same metrics collection approach explained in Appendix~\ref{memory-metrics-collection-approach}. 

Overall, ESBMC v7.3 with Boolector is the fastest configuration that also consumes the minimum amount of memory to verify all benchmarks. Among the other solvers, the memory consumption of ESBMC v7.3 with Bitwuzla comes near the Boolector configuration. This is probably because Bitwuzla is an extended forked Boolector~\cite{niemetz2020bitwuzla}. We also found that MathSAT tends to use more memory for timed-out test cases, as ESBMC v7.3 with MathSAT failed only one TC in the $cpp\-sub$ benchmark due to timeout but uses more memory than the other configurations.

\begin{table}[h]
\caption{Experimental results showing the pass rate and total verification time for ESBMC using different solvers.}
\begin{center} 
\begin{tabular}{|c|c|c|c|c|c|c|c|c|c|c|c|}
\hline % \cline{i-j}, line from column i to column j
\textbf{Sub-Benchmarks} &\textbf{Boolector} &\textbf{CVC4} &\textbf{MathSAT} &\textbf{Yices} &\textbf{Z3} &\textbf{Bitwuzla}\\ \hline
    cpp-sub                 &{$100\%$} &{$99\%$}  &{$99\%$}  &{$100\%$} &{$100\%$} &{$100\%$}\\\hline
    inheritance-sub         &{$100\%$} &{$93\%$}  &{$100\%$} &{$100\%$} &{$100\%$} &{$100\%$}\\\hline
    polymorphism-sub        &{$100\%$} &{$100\%$} &{$100\%$} &{$100\%$} &{$100\%$} &{$100\%$}\\\hline
    cbmc-template-sub       &{$100\%$} &{$97\%$}  &{$100\%$} &{$100\%$} &{$100\%$} &{$100\%$}\\\hline
    gcc-template-tests-sub  &{$100\%$} &{$96\%$}  &{$100\%$} &{$100\%$} &{$100\%$} &{$100\%$}\\\hline
    template-sub            &{$100\%$} &{$92\%$}  &{$100\%$} &{$100\%$} &{$100\%$} &{$100\%$}\\\hline
\textbf{Total verification Time} &{$128.796s$} &{$637.988s$} &{$131.934s$} &{$182.327s$} &{$162.848s$} &{$152.442$}\\\hline
\end{tabular}
\label{table:results-solvers-time}
\end{center}
\end{table}

\begin{table}[h]
\caption{Experimental results showing the memory usage for ESBMC using different solvers.}
\begin{center} 
\begin{tabular}{|c|c|c|c|c|c|c|c|c|c|c|c|}
\hline % \cline{i-j}, line from column i to column j
\textbf{Sub-Benchmarks} &\textbf{Boolector} &\textbf{CVC4} &\textbf{MathSAT} &\textbf{Yices} &\textbf{Z3} &\textbf{Bitwuzla}\\ \hline
    cpp-sub                 &{$19385$ MB} &{$63757$ MB} &{$153326$ MB}  &{$27983$ MB} &{$35758$ MB} &{$19455$ MB}\\\hline
    inheritance-sub         &{$845$ MB} &{$950$ MB}   &{$940$ MB} &{$847$ MB} &{$946$ MB} &{$855$ MB}\\\hline
    polymorphism-sub        &{$2373$ MB} &{$2657$ MB} &{$2632$ MB} &{$2320$ MB} &{$2596$ MB} &{$2387$ MB}\\\hline
    cbmc-template-sub       &{$2295$ MB} &{$2558$ MB} &{$2449$ MB} &{$2308$ MB} &{$2457$ MB} &{$2299$ MB}\\\hline
    gcc-template-tests-sub  &{$1387$ MB} &{$1559$ MB} &{$1480$ MB} &{$1401$ MB} &{$1497$ MB} &{$1395$ MB}\\\hline
    template-sub            &{$727$ MB} &{$800$ MB}   &{$781$ MB} &{$730$ MB} &{$774$ MB} &{$733$ MB}\\\hline 
\textbf{Total memory} &{$27012$ MB} &{$72281$ MB} &{$161608$ MB} &{$35589$ MB} &{$44028$ MB} &{$27124$ MB}\\\hline
\end{tabular}
\label{table:results-solvers-memory}
\end{center}
\end{table}

%====================================================================
\section{Planning for Future work}
\label{planning}
%====================================================================
ESBMC v2.1 mimics the semantics of the APIs of C++ STL libraries using a set of operational models (OMs). The C++ front-end of ESBMC has been completely rewritten, and the back-end has also undergone significant development and evolution since v2.1 was published in~\cite{monteiro2022model}, therefore it is questionable whether those OMs still work. We believe that it is essential to evaluate both ESBMC v2.1 and v7.3 with the existing OMs over the C++ library benchmarks from~\cite{monteiro2022model}. Table~\ref{table:results-other-categories} provides a summary of the pass rates. 

As shown in Table~\ref{table:results-other-categories}, the OMs of ESBMC v2.1 give a fairly good pass rate of $80\%$ and more in most of the benchmarks, except for STL $algorithm$, $list$, $multiset$ and $vector$, where the pass rate is below $50\%$. In ESBMC, there are a total of 63 C++ OMs representing the most frequently used STL libraries. 41 out of 63 OMs have been enabled with the new clang C++ front-end. The remaining OMs are still being fixed. As we continue to enable more OMs, the progress is tracked and publicly available for viewing on ESBMC's wiki page \textit{OM Workload Estimate and Tracking}~\footnote{\url{https://github.com/esbmc/esbmc/wiki/OM-Workload-Estimate-and-Tracking}}. 

\begin{table}[h]
\caption{Pass rates of OM-dependent benchmarks for C++ STL libraries.}
\begin{center} 
\begin{tabular}{|c|c|c|c|c|c|c|c|c|c|c|c|}
\hline % \cline{i-j}, line from column i to column j
\textbf{Benchmarks} &\textbf{ESBMC-v2.1 pass rate} &\textbf{ESBMC post-v7.3 pass rate} \\ \hline
    string                 &{$99\%$}       &{$0\%$}     \\\hline
    stream                 &{$89\%$}      &{$33\%$}     \\\hline
    algorithm              &{$42\%$}       &{$0\%$}     \\\hline
    deque                  &{$95\%$}       &{$0\%$}     \\\hline
    list                   &{$53\%$}       &{$0\%$}     \\\hline
    map                    &{$83\%$}       &{$0\%$}     \\\hline
    multimap               &{$89\%$}       &{$0\%$}     \\\hline
    multiset               &{$74\%$}       &{$0\%$}     \\\hline
    priority\_queue        &{$100\%$}      &{$0\%$}     \\\hline
    set                    &{$83\%$}       &{$0\%$}     \\\hline
    stack                  &{$86\%$}       &{$0\%$}     \\\hline
    vector                 &{$22\%$}       &{$0\%$}     \\\hline
    try\_catch             &{$88\%$}       &{$0\%$}     \\\hline

\end{tabular}
\label{table:results-other-categories}
\end{center}
\end{table}

\end{document}